\documentclass[conference,a4paper]{IEEEtran}

\usepackage{xcolor}
\usepackage{balance}
\ifCLASSINFOpdf
  \usepackage[pdftex]{graphicx}
\else
  \usepackage[dvips]{graphicx}
\fi
\usepackage{cite}
\usepackage{epstopdf}
\usepackage{multirow}
\usepackage{siunitx}
\usepackage{amsmath}
\usepackage{siunitx}
\usepackage{pgfplots}
\pgfplotsset{compat=newest}
\usepackage{nicefrac}
\newlength\figureheight
\newlength\figurewidth
\newcommand{\figref}{Fig.~\ref}
\newcommand{\tabref}{Table~\ref}

\usepackage{comment}
\ifCLASSOPTIONcompsoc
 \usepackage[caption=false,font=normalsize,labelfont=sf,textfont=sf]{subfig}
\else
 \usepackage[caption=false,font=footnotesize]{subfig}
\fi
\usepackage[absolute,overlay]{textpos} 
\usepackage{array}
\newcolumntype{B}[1]{>{\centering\arraybackslash}b{#1}}

\begin{document}

\title{Energy-Efficient Physical Layer Security for Wearable IoT Devices}

\author{\IEEEauthorblockN{
Abel Zandamela\IEEEauthorrefmark{1},   
Nicola Marchetti\IEEEauthorrefmark{1},   
Adam Narbudowicz\IEEEauthorrefmark{1}    
}                                     
\IEEEauthorblockA{\IEEEauthorrefmark{1}%
CONNECT Centre, Trinity College Dublin, The University of Dublin, Dublin, Ireland.\\ \{zandamea, nicola.marchetti, narbudoa\}@tcd.ie}
}

\maketitle

\begin{abstract}
This work proposes an energy-efficient Directional Modulation (DM) scheme for on-body Internet of Things (IoT) devices. DM performance is tested using a 5-port stacked-patch MIMO antenna under two scenarios: free space case and using a four-layer human forearm phantom to simulate the user's wrist. It is demonstrated that the scheme achieves steerable secure transmissions across the entire horizontal plane. With low Bit Error Rate (BER) of $\boldsymbol{1.5\times10^{-5}}$ at the desired directions, eavesdroppers experience a high error rate of up to $\boldsymbol{0.498}$. Furthermore, this work investigates the DM performance using a subset of the stacked patches in the MIMO antenna, revealing that some combinations achieve a low BER performance using a lower antenna profile, albeit high side-lobes of BER$\boldsymbol{<10^{-2}}$ seen outside the desired region. Overall, the solution is proposed as a good candidate to enable secure wireless communications in emerging wearable IoT devices that are subject to size and energy-constraints.
\end{abstract}

\vskip0.5\baselineskip
\begin{IEEEkeywords}
 Physical layer security (PLS), directional modulation (DM), MIMO antennas, wearable IoT devices, compact IoT devices, on-body IoT systems, pattern reconfigurable antennas.
\end{IEEEkeywords}

\section{Introduction}

The Internet of Things (IoT) technology is increasingly shaping our daily-life, making IoT devices key enablers of cutting-edge technologies like smart cities and healthcare \cite{Qadri2020, Dian2020}. On-body IoT devices such as wrist-worn devices are becoming indispensable for many modern applications, e.g. patient monitoring, in-hospital localization and tracking, sports fitness and tracking, smart building monitoring \cite{Seneviratne2017, Dian2020}. However, miniaturization, multi-functionality and low energy-consumption capabilities are key design challenges in wearable IoT devices; such characteristics limit implementation of more traditional security methods for wearable IoT scenarios \cite{Seneviratne2017}. 

Physical Layer Security (PLS) is an emerging technique proposed to enhance wireless transmissions by exploiting the physical characteristics of the wireless channel. An example of PLS solution is the Directional Modulation (DM) method, which transmits intelligible modulation only in the direction of the intended legitimate receiver \cite{Trung2018, Wei2020}. In state-of-the-art it is executed using antenna arrays \cite{Daly2009,Ding2014,Shu2018}, which due to the inter-element spacing, are too large for wearable IoT devices at sub-$\SI{6}{GHz}$ bands. DM implementations using small antennas are discussed in \cite{Adam2017,Hernandez2021, EuCAP2022}. However, those implementations require simultaneous port activation to realize DM, which is still challenging for wearable IoT systems, as the necessary number of active RF transceivers increases the cost and exacerbates the tight energy constraint of wearable technologies. Single port DM implementations using time-modulated arrays are proposed in \cite{Huang2021,Huang2022}; however, these methods are still demanding for wearable IoT implementation due to switch synchronization (which requires more computational power), and larger space for antenna packaging. In our previous work \cite{AWPL2022}, we propose a single-port DM method relying on switches but without the need for any specific sequence, i.e. the switching can be controlled by a fully random mechanism, independent from the transceiver. However, the method was demonstrated using a 5-element circular array which still has too large a profile and size for wearable devices.  

In this work, we propose a single-port DM scheme for on-body IoT devices. The system is tested using a four-layer human forearm phantom and a 5-port stacked patch antenna capable of unidirectional beamsteering around the entire horizontal plane \cite{EuCAP2022}. It is demonstrated that a low Bit Error Rate (BER) of $1.5\times10^{-5}$ is experienced by the desired receiver, while a high error rate of up to $0.498$ is observed outside the desired regions. In addition, to allow for advanced DM properties like multi-target transmissions, subsets comprising different combinations of stacked-patches of the investigated antenna are used for feasibility of multiport DM with reduced number of RF transceivers. It is demonstrated that DM transmissions can be realized using a lower profile device, reduced number of ports and higher efficiency, albeit for such a case the compromise requires side-lobes in some directions.

\section{Simulation Setup}

\subsection{Antenna Configuration}

\begin{figure}
\centering
\subfloat[]{\includegraphics[clip, trim=0cm .9cm 0cm 0cm,width=1.0\columnwidth]{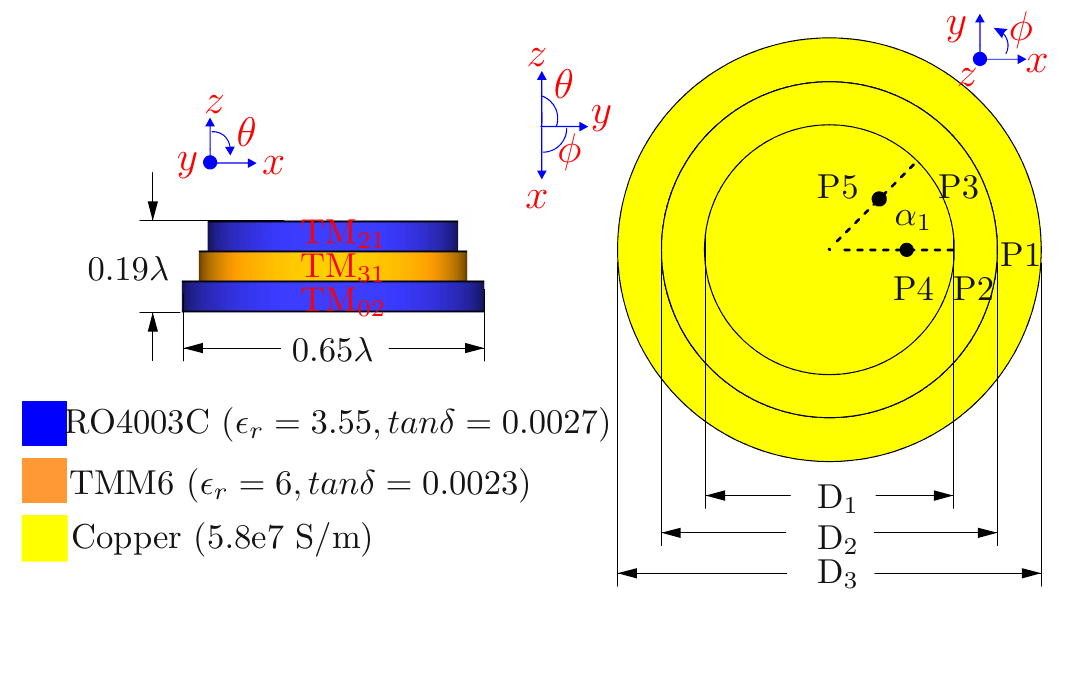}%
\label{fig:invest_antenna}}
\hfil
\subfloat[]{\includegraphics[clip, trim=0cm 2cm 3cm 0.5cm, width=.5\columnwidth]{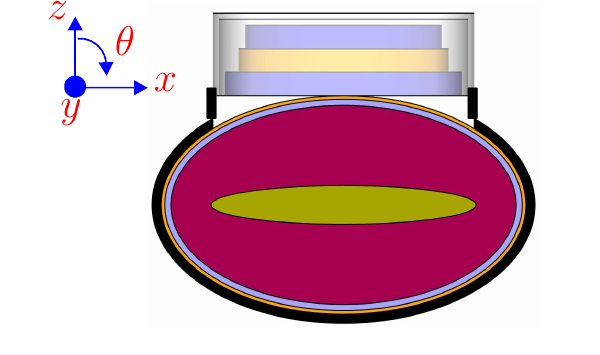}%
\label{fig:phantom}}
\caption{Simulation setup to investigate the performance of the energy-efficient DM in the user's wrist scenario: (a) front and top views of the 5-port stacked patch antenna \cite{EuCAP2022}; (b) antenna integrated into a plastic housing in direct contact with a four-layer human forearm phantom.}
\label{fig:system_setup}
\end{figure}

\begin{figure}[!htbp]
\centering
{\includegraphics[width=1.0\columnwidth]{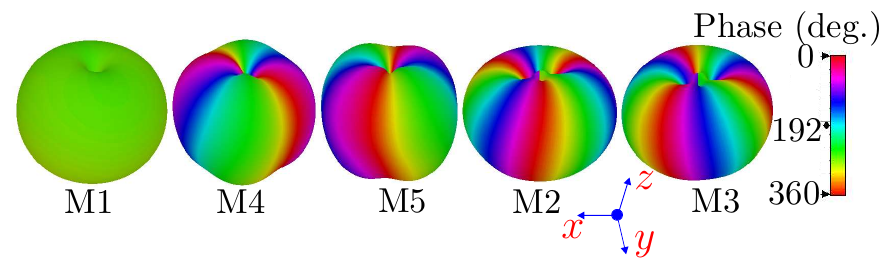}}
\caption{3D phase patterns of all the modes excited by the antenna in \figref{fig:invest_antenna}. Note that each mode shows different phase variations around the horizontal plane; this key property is exploited to enable beamforming characteristics within a compact-wearable antenna.}
\label{fig:ant_phases}
\end{figure}

To analyze the performance of the energy-efficient directional modulation scheme, a simulation setup comprising a multimodal beamsteering stacked-patch antenna proposed in \cite{EuCAP2022} and a forearm phantom are used. For completeness, the design is shown in \figref{fig:invest_antenna}; it comprises three stacked dielectric-loaded patches designed to operate at the center frequency of $\SI{5}{GHz}$ and fed using 5-ports to excite five different orthogonal modes. The top patch has diameter $D_1=\SI{30.4}{mm}$ and is fed using P4 and P5 rotated by $\alpha_1=\ang{45}$; such arrangement excites two orthogonal TM$_{21}$ modes. The middle patch has diameter $D_2=\SI{34.6}{mm}$ and is fed using P2 and P3 rotated by $\ang{30}$ to excite two orthogonal TM$_{31}$ modes. Finally, the bottom patch has the largest diameter, $D_3 = \SI{78}{mm}$, and is fed using P1 to generate a monopole-like TM$_{02}$ mode. The ports of the top and middle patches are fed with $\pm\ang{90}$ phase shift between each respective patch pair, to ensure electric field phase rotation around their perimeters. In this manner the obtained modes (shown in \figref{fig:ant_phases}), have a constant phase for M1, dual phase-change in opposite directions for M4 and M5, and triple phase-change also in opposite directions for M2 and M3. To achieve unidirectional beamsteering around the horizontal plane, the phase variations of the excited modes are exploited to create constructive interference in the desired directions around the same plane ($xy-$plane).    
To evaluate the antenna performance on the user's wrist, the antenna is placed inside a plastic housing and a four-layer human forearm phantom (comprising skin, fat, muscle and bone layers) of $\SI{200}{mm}$ and width~$=\SI{60}{mm}$ is used. In all the following results discussion, the scenario to evaluate the user's wrist, assumes that the antenna is placed against the skin layer (outer-most layer) of the phantom as shown in \figref{fig:phantom}.

\subsection{Signal Processing}

The Directional Modulation (DM) scheme is proposed to enable secure wireless transmissions in energy-constrained wearable IoT systems. The scheme relies on switches in a multiport antenna structure, requiring only a single active port for transmission. It then uses phase compensations dependent on the desired secure direction of the legitimate user, and a random number $n$ corresponding to the antenna port used for signal transmission (see \cite{AWPL2022} for a detailed discussion of the scheme). The method then offers several advantages compared to \cite{Hernandez2021, EuCAP2022}, e.g.: no excessive artificial noise is generated in the directions where the illegitimate users are, which reduces interference with legitimate users present in other systems. Additionally, because only a single port activation is needed at a time, the scheme allows for energy-saving when compared to schemes with simultaneous RF transceivers activation.

\begin{figure}[!htbp]
\centering
{\includegraphics[clip, trim=0cm 15cm 4cm 0cm, width=1.0\columnwidth]{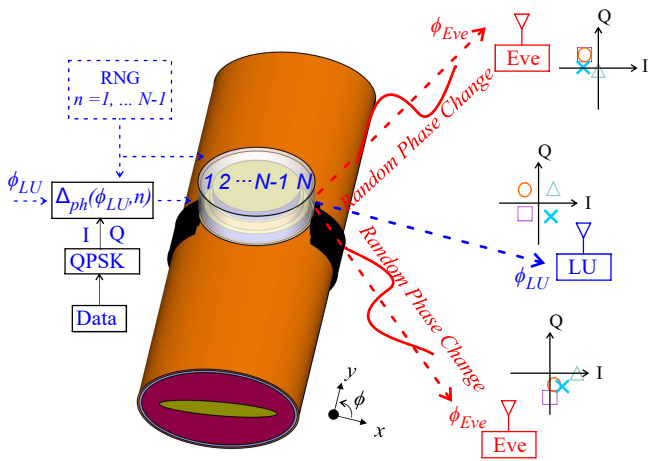}}
\caption{Schematic outlining the DM implementation for on-body wrist-wearable IoT devices. From the left it is seen that the DM symbol transmission is executed using QPSK with phase compensation dependent on the direction of the Legitimate User (LU) and the value of the Random Number Generator (RNG). The method allows for distorted IQ constellations in all directions except for $\phi_{LU}$ direction, as seen on the right side of the image.}
\label{fig:dirmod_scheme}
\end{figure}

The proposed implementation for on-body IoT devices is shown in \figref{fig:dirmod_scheme}. Without loss of generality the scheme is evaluated using Quadrature Phase Shift Keying (QPSK) modulation, and the direction of the Legitimate User (LU) is denoted $\phi_{LU}$. In the initial stage a Random Number Generator (RNG) is used for selection of the reference port. Next the symbols to be transmitted using QPSK go through a phase delay $\Delta_{ph}(\phi_{LU},n)$; this allows to compensate for the phase variations when switching between the remaining $N-1$ ports of the wrist-wearable IoT device, while simultaneously arbitrary phase changes are observed for receivers located outside the desired secure direction. This makes it challenging for the eavesdroppers to retrieve the transmitted data and is highlighted in \figref{fig:dirmod_scheme}, where the DM transmissions can be performed for any angle in the horizontal plane ($xy-$plane), and is seen that the In-phase and Quadrature (IQ) constellations are only decipherable at the desired $\phi_{LU}$ direction, while distorted IQ patterns are observed in all other directions. To verify the feasibility of the energy-efficient DM scheme for wearable devices, numerical studies were conducted, with Bit Error Rate (BER) calculations executed using $10^5$ transmitted symbols, $\SI{12}{dB}$ Signal to Noise Ratio (SNR), and Additive White Gaussian Noise (AWGN) added in the receiver and assumed to be independent in each direction. 

\section{Simulations Results}

This Section presents the numerical results of the proposed setups. The farfield results are obtained using CST Studio Suite finite element method and the directional modulation implementation is executed in MATLAB.

\subsection{Energy-Efficient DM Performance}

\begin{figure}[htbp]
    \centering
	\subfloat[]{
	\setlength\figureheight{0.22\textwidth}
	\setlength\figurewidth{0.37\textwidth}
	\input{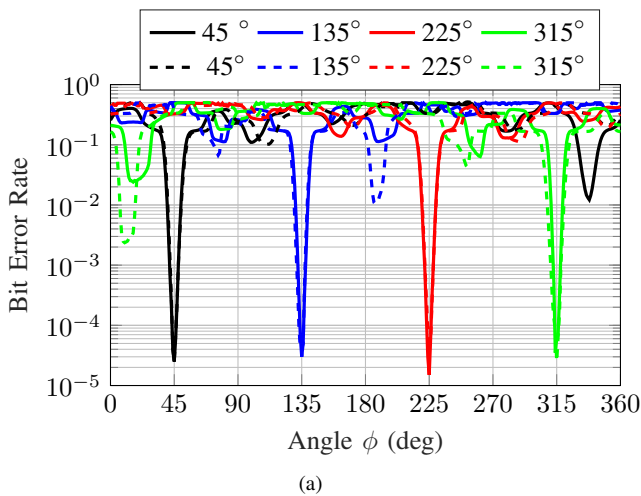}}
	\caption{BER calculations of the proposed single-active port energy-efficient DM using $\SI{12}{dB}$ SNR for the antenna in free space (solid lines) and including the forearm phantom (dashed lines) for different $\phi_{LU}$ directions.}
\label{fig:DM_energyefficient}
\end{figure}

\begin{figure}[ht]
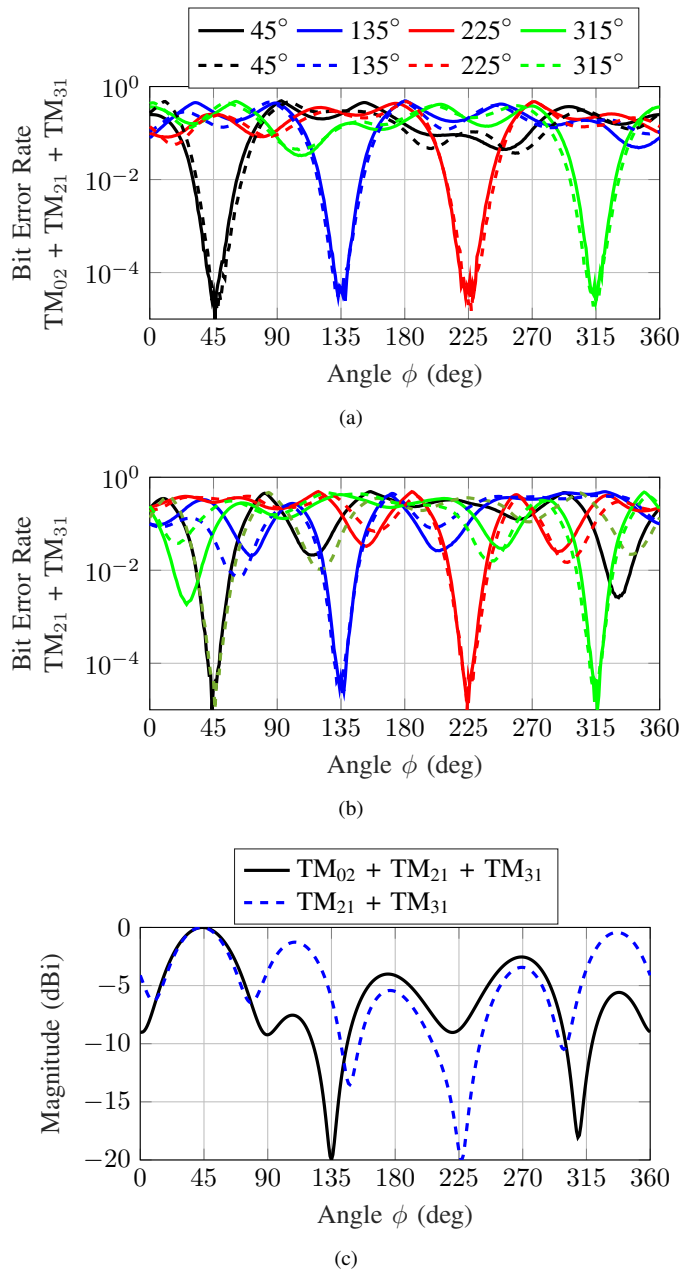

    \centering
	\subfloat[]{
    \hspace{-0.5cm}
	\setlength\figureheight{0.17\textwidth}
	\setlength\figurewidth{0.37\textwidth}
	\input{NewCompar3Modes}
	\label{fig:multi3modes}}
	\hfill
	\subfloat[]{
	\hspace{-0.5cm}
	\setlength\figureheight{0.17\textwidth}
	\setlength\figurewidth{0.37\textwidth}
	\input{NewCompar2Modes}
	\label{fig:multi2modes}}
	\hfill
	\subfloat[]{
	\setlength\figureheight{0.17\textwidth}
	\setlength\figurewidth{0.37\textwidth}
	\input{5P_4P_Beam_Compar}
	\label{fig:beam_compar_p4_p5}}
	\caption{BER calculations for the simultaneous multiport activation DM \cite{EuCAP2022}, using $\SI{12}{dB}$ SNR for the antenna in free space (solid lines) and including the forearm phantom (dashed lines) for different $\phi_{LU}$ directions to cover the entire horizontal plane: (a) DM implementation for $N=5$ ports/modes; (b) DM implemented for a subset formed only by P2, P3, P4 and P5 ($N=4$); and (c) free space normalized full-wave simulated patterns comparing the beamsteering performance between $N=5$ ports/modes and $N=4$ subset for $\phi=\ang{45}$.}
\label{fig:DM_compar}
\end{figure}

\figref{fig:DM_energyefficient} shows the DM performance for free space case (shown in solid lines) and including the forearm phantom (shown in dashed lines). The configuration assumes that all $N=5$ modes/ports are available for DM transmissions; therefore, after the selection of the reference port for handshake, any of the $N-1$ remaining ports can be randomly selected for data transmission. The performance is investigated for four different legitimate user directions separated by $\ang{90}$ to scatter across the entire horizontal plane: $\phi_{LU}=\ang{45}$, $\ang{135}$, $\ang{225}$, and $\ang{315}$. It can be seen that in both scenarios (free space and user's wrist) a unique secure steerable transmission is achieved without leakage into undesired directions for all the four cases. In detail, in both scenarios the LU experiences a very low bit error rate (BER) of $1.5\times10^{-5}$, while a high error rate of $0.498$ is observed outside the desired region. Moreover, low beamwidths of $\ang{10}$ (for BER$<10^{-2}$) are observed in all the four investigated transmission directions. It should be noted however, that for some directions a second BER$<10^{-2}$ region is observed when the phantom effects are included. This maybe attributed to pattern tilts due to the phantom for those specific regions. Overall, the achieved results show a very promising solution to enhance the communications of wearable IoT devices, as such low BER performance is realized using a single active port when compared to the method presented in \cite{EuCAP2022}.

\subsection{Energy Reduction Analysis for Multiport DM}

Because the use of simultaneous multiport activation DM schemes offer additional advantages such as multi-target DM transmissions, we also investigate the feasibility of reducing the number of ports required for DM transmissions on the method discussed in \cite{EuCAP2022}. The advantages is that multiple simultaneous secure transmissions can be realized while also allowing for lower energy consumption in resource-constrained IoT devices. As the antenna consists of stacked patches exciting different orthogonal radiating modes, subsets formed by combinations of the three collocated patches are analyzed to reduce the number of RF transceivers.

\figref{fig:DM_compar} shows the multiport activation DM performance in free space and including the phantom when all five ports of the antenna are active (\figref{fig:multi2modes}) and for the subset comprising the top (TM$_{21}$ modes, P4 and P5) and middle (TM$_{31}$ modes, P2 and P3) patches (\figref{fig:multi2modes}). The results highlight that in both cases the the lowest BER is seen at the desired $\phi_{LU}$ directions. For the case using $N=5$ active ports a BER of $10^{-5}$ is realized at $\phi_{LU}$ and the beamwidth with BER$<10^{-2}$ is around $\ang{32}$. The (TM$_{21}$ $+$ TM$_{31}$) subset using $N=4$ ports also achieves BER of $10^{-5}$. However, outside the desired directions a second region with BER$<10^{-2}$ is also observed for $\phi_{LU} = \ang{45}$ and $\ang{315}$ cases. Even though these sidelobes are smaller for the phantom case, they can be exploited by eavesdroppers to partially retrieve the transmitted data; on the other hand, its BER$<10^{-2}$ beamwidth in the $\phi_{LU}$ directions is narrower (around $\ang{24}$). It should also be noted that the high sidelobes of the (TM$_{21}$ + TM$_{31}$) subset are due to the beamsteering performance of the configuration; this is seen in \figref{fig:beam_compar_p4_p5}, where the beamsteering of the $N=5$ configuration (using all 5 modes/ports) achieves lower sidelobes compared to the $N=4$ subset (without the TM$_{02}$) mode. Even though the mainbeam is at the desired $\phi=\ang{45}$ for $N=4$ subset, it can be seen that two sidelobes are also present around $\ang{115}$ and $\ang{335}$. Overall, for a system comprised only by this subset (TM$_{21}$ $+$ TM$_{31}$), energy saving is possible by reducing the number of active RF transceivers from $N=5$ to $N=4$. In addition the profile of the antenna can be reduced to $0.12\lambda$, which can lower the height of the IoT device.

\begin{figure}[ht]
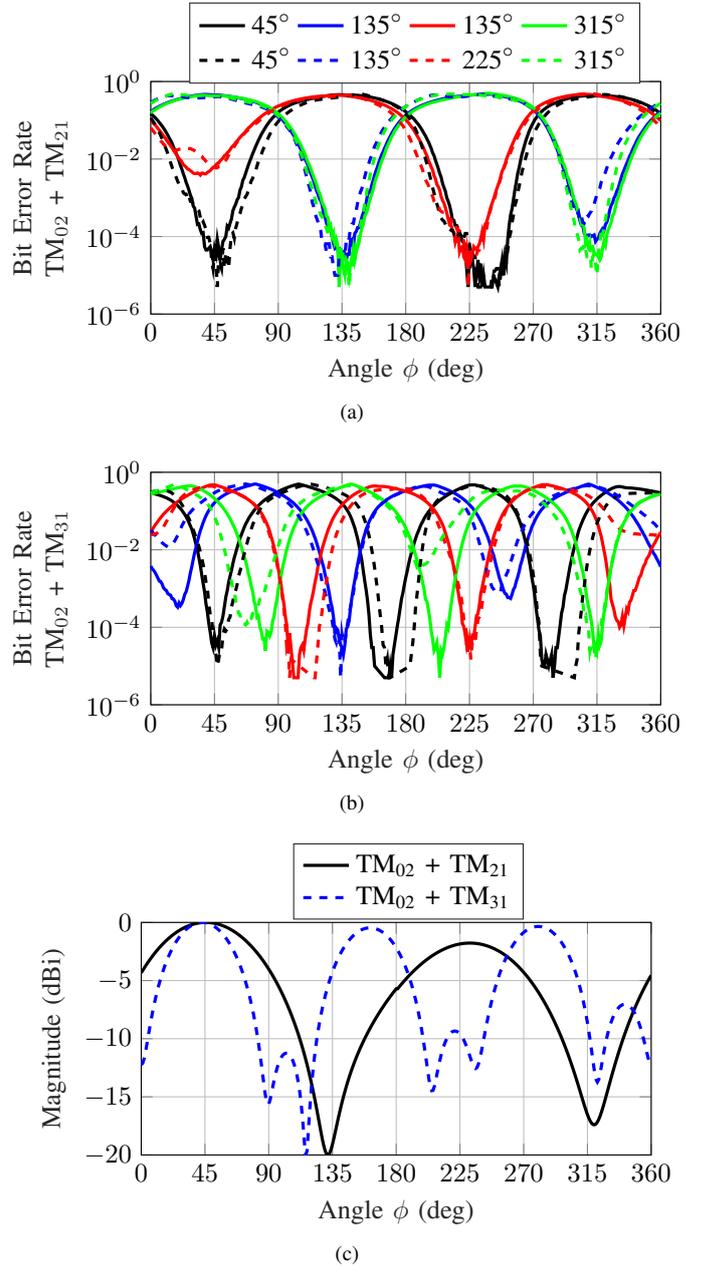

	\subfloat[]{
	\hspace{-0.5cm}
	\setlength\figureheight{0.17\textwidth}
	\setlength\figurewidth{0.37\textwidth}
	\input{NewCompar2ModesTM21}
	\label{fig:dirmod_3PTM21}}
	\hfill
	\subfloat[]{
	\hspace{-0.5cm}
	\setlength\figureheight{0.17\textwidth}
	\setlength\figurewidth{0.37\textwidth}
	\input{NewCompar2ModesTM31}
	\label{fig:dirmod_3PTM31}}
	\hfill
	\subfloat[]{
	\setlength\figureheight{0.17\textwidth}
	\setlength\figurewidth{0.37\textwidth}
	\input{3P_Beam_Compar}
	\label{fig:beam_compar_p3}}
	\caption{BER calculations using $\SI{12}{dB}$ SNR for the antenna in free space (solid lines) and including the forearm phantom (dashed lines) for different $\phi_{LU}$ directions to cover the entire horizontal plane: (a) DM implementation using TM$_{02}$ and TM$_{21}$ combination; (b) DM implementation for a subset formed only by the ports exciting TM$_{02}$ and TM$_{31}$ modes; and (c) free space normalized full-wave simulated patterns comparing the beamsteering performance between (TM$_{02}$ + TM$_{21}$) configuration and the (TM$_{21}$ + TM$_{31}$) subset for $\phi=\ang{45}$.}
    \label{fig:DM_compar2}
\end{figure}

\begin{table*}[!htbp]
\caption{Comparisons of the DM performance for different investigated antenna configurations.}
\label{tab:DM_compar}
\centering
\centering
\begin{tabular}{|B{2cm}|B{2cm}|B{1.5cm}|B{1.5cm}|B{1.5cm}|B{1.5cm}|B{1.6cm}|}
\hline
Antenna Configuration                     & Beam Pattern                                                             & BER at $\phi_{LU}$ & \begin{tabular}[c]{@{}c@{}}Regions with\\  BER$<10^{-2}$\end{tabular} & BER$<10^{-2}$ & Antenna Profile & \begin{tabular}[c]{@{}c@{}}Total efficiency\\ ($\SI{5}{GHz}$)\end{tabular} \\ \hline
P1, P2, P3, P4, P5 & \begin{tabular}[c]{@{}c@{}}Unidirectional\\ Low side-lobes\end{tabular}  & $10^{-5}$   & 1                                                                     & $\ang{32}$    & $0.19\lambda$   & $49\%$                                                                                          \\ \hline
P2 P3, P4, P5      & \begin{tabular}[c]{@{}c@{}}Unidirectional\\ High side-lobes\end{tabular} & $1.5\times10^{-5}$ & 2                                                                     & $\ang{24}$    & $0.12\lambda$   & $49\%$                                                                                          \\ \hline
P1, P4, P5       & Bi-directional                                                           & $2.5\times10^{-5}$ & 2                                                                     & $\ang{51}$    & $0.12\lambda$   & $87\%$                                                                                          \\ \hline
P1, P2, P3       & Tri-directional                                                          & $3.5\times10^{-5}$ & 3                                                                     & \ang{27}             & $0.12\lambda$   & $49\%$                                                                                          \\ \hline
\end{tabular}
\end{table*}

\figref{fig:DM_compar2} shows the performance of the subsets comprising the bottom and top patches (TM$_{02}$ + TM$_{21}$) or (P1, P4 and P5) illustrated on \figref{fig:dirmod_3PTM21} and the bottom and middle patches (TM$_{02}$ + TM$_{31}$) or (P1, P2 and P3) seen in \figref{fig:dirmod_3PTM31}, for both the antenna in free space and including the forearm phantom. First, for the TM$_{02}$ + TM$_{21}$  (\figref{fig:dirmod_3PTM21}) configuration, the results show that it produces two directions with BER smaller than $10^{-2}$. Note that one of the directions correspond to the desired $\phi_{LU}$ angle, while the second one is seen at $\phi_{LU} + \ang{180}$. This is explained by the beamsteering performance of the subset; as shown in \figref{fig:beam_compar_p3} the configuration produces a bi-directional pattern with the two mainbeams separated by $\ang{180}$. Nevertheless, within the $[\ang{0} - \ang{180}]$ region, the subset shows only one low BER region (around $2.5\times10^{-5}$), which is at the desired $\phi_{LU}$ direction; outside of this region BER values of $0.475$ are observed for undesired users. This subset also allows for antenna profile reduction to $0.12\lambda$, with improved efficiency and bandwidth. This is because the lowest efficiency and the overlapping bandwidth for the three stacked-patches (including all $N=5$ ports/modes of the antenna) is limited by the performance of the higher-order TM$_{31}$ modes of the middle patch \cite{EuCAP2022}. Since, the TM$_{31}$ modes are not included in the above subset, performance improvements in terms of efficiency and bandwidth are realized, albeit the ambiguities in the DM performance (see \figref{fig:dirmod_3PTM21}).

\figref{fig:dirmod_3PTM31} shows the performance of the subset formed by the TM$_{02}$ and TM$_{31}$ modes. It is seen that this subset has the worst performance, because it produces three regions with BER$<10^{-2}$ in all the four investigated $\phi_{LU}$ directions, for both scenarios: antenna in free space and with the human forearm phantom. Low BER performance is also seen at the desired receiver ($\phi_{LU}$) with values around $3.5\times10^{-5}$; however two other directions with low BER are also present during the signal transmission and are located at $\phi_{LU} + \ang{120}$ and $\phi_{LU} + 2\times\ang{120}$ directions. Note that these angles are related to the beamsteering performance of the configuration; as depicted in  \figref{fig:beam_compar_p3}, compared to the subset comprising (TM$_{02}$ + TM$_{21}$), this configuration produces three mainbeams across the entire horizontal plane.

Lastly, \tabref{tab:DM_compar} shows the multiport DM performance comparisons between the different configurations. Overall, it is seen that the configuration using all the $N=5$ modes/ports offers a DM performance with low side-lobes levels and the lowest BER values of $10^{-5}$; however it requires an antenna profile of $0.19\lambda$ and has $49\%$ total antenna efficiency. With a trade-off on the number of regions with BER$<10^{-2}$ a lower antenna profile is feasible using using $N=4$ and $N=3$ configurations. The $N=4$ subset offers better side lobes levels and BER when compared to $N=3$, however, for applications requiring high antenna efficiency the $N=3$ (TM$_{02}$ + TM$_{21}$) configuration offers better performance with total efficiency of $87\%$ and $0.12\lambda$ antenna profile. 

\section{Conclusion}
This work uses orthogonal radiating modes excited in a stacked-patch configuration to achieve unidirectional beamsteering across the entire horizontal plane. This performance is exploited to realize directional modulation transmissions requiring only a single-port activation at a time. It was demonstrated that low BER performance of $1.5\times10^{-5}$ is achieved in the direction of the legitimate user, without leakage into other undesired directions. Moreover, the work demonstrated the use of different orthogonal modes to study the feasibility of reduced number of RF transceivers in a multiport activation DM system. The solution illustrated that albeit a side-lobe with BER$<10^{-2}$, a steerable secure transmission can be realized using a combination of TM$_{02}$ and TM$_{21}$ modes, with   $87\%$ efficiency and a profile of $0.12\lambda$. 

\section*{Acknowledgment}
This publication has emanated from research conducted with the financial support of Science Foundation Ireland under Grant number $18$/SIRG/$5612$ and in part by the IEEE Antennas and Propagation Society Doctoral Research Grant 2022.

\balance

\end{document}